\documentstyle[aps,prl,epsf,preprint]{revtex}
\def\fun#1#2{\lower3.6pt\vbox{\baselineskip0pt\lineskip.9pt
        
\ialign{$\mathsurround=0pt#1\hfill##\hfil$\crcr#2\crcr\sim\crcr}}}
\begin{document}
\title{\vskip-2.5truecm{\hfill \baselineskip 14pt 
{\hfill {{\small        \hfill BA-97-25}}}\\
{{\small        \hfill CERN-TH/97-275}}\vskip .1truecm} 
\vspace{1.0cm}
\vskip 0.1truecm {\bf $\mu$ Problem and Hybrid Inflation 
in Supersymmetric ${\bf SU(2)_L \times SU(2)_R \times 
U(1)_{B-L}}$\thanks{To be submitted to Phys. Lett. {\bf B}}}}
\author{{G. Dvali}$^{(1)}$, {G. Lazarides}$^{(1,2)}$ {and Q. 
Shafi}$^{(3)}$} 
\address{$^{(1)}${\it Theory Division, CERN, CH-1211, Geneva 23, 
Switzerland}}
\address{$^{(2)}${\it Physics Division, School of Technology, Aristotle 
University of Thessaloniki,\\ Thessaloniki GR 540 06, Greece.}}
\address{$^{(3)}${\it Bartol Research Institute, University of 
Delaware,\\
Newark, DE 19716, USA}} 
\maketitle


\begin{abstract}
\baselineskip 12pt
We present a solution of the $\mu$ problem within a supersymmetric model 
based on the 
$SU(2)_L\times SU(2)_R\times U(1)_{B-L}$ gauge group. We show that this 
solution implies a built-in hybrid inflationary scenario, which may 
result in a
successful baryogenesis via leptogenesis.  Baryon number is essentially 
conserved as a result of an R-symmetry.  Neutrinos with mass in 
the eV range can provide the `hot' dark matter component.
\end{abstract}
\thispagestyle{empty}
\newpage
\pagestyle{plain}
\setcounter{page}{1}
\def\beq{\begin{equation}}
\def\eeq{\end{equation}}
\def\beqa{\begin{eqnarray}}
\def\eeqa{\end{eqnarray}}
\def\tr{{\rm tr}}
\def\x{{\bf x}}
\def\p{{\bf p}}
\def\k{{\bf k}}
\def\z{{\bf z}}
\baselineskip 20pt

It is well known that the minimal supersymmetric standard model (MSSM)
leaves unanswered a number of fundamental questions. For instance:
$i$) How does the well known parameter $\mu$ originate?
$ii$) How is the inflationary scenario realized?
$iii$) How is `hot' dark matter (in neutrinos with mass in the eV 
range) incorporated in order
to explain the observed large scale structure formation, especially if 
the
primordial density fluctuations are (essentially) scale invariant?
$iv$)  Why is the proton so stable, even though `matter' parity allows 
the dimension five process?

 In this paper, we will investigate an extension of MSSM
which provides a resolution of these issues.
In our analysis, the new physics
enters at a scale $M$ well above the weak scale $M_Z$ so that,
when the superheavy states are integrated out, we obtain MSSM
with the $\mu$ parameter determined.
It turns out, as we will see, that a natural candidate for such a 
structure
is based on the gauge group $G_{LR} = SU(3)_c\times SU(2)_R\times 
SU(2)_L \times U(1)_{B-L}$. Left-right (L-R) symmetric models based on
this gauge group have been extensively studied in the past\cite{lr},
as well as more recently \cite{susylr} in the supersymmetric context. 
Since
our focus lies on the points listed above we shall, for simplicity, 
relax the requirement of L-R symmetry
in the higgs sector and concentrate on the minimal possible version of a 
theory based on the gauge group
$G_{LR}$.  We later indicate how this scheme
could be embedded in a fully L-R symmetric model
which, in principle, may be the complete $SO(10)$ model.
 
Before proceeding, let us briefly summarize our results here and also 
try to
justify our choice of the gauge group.
Our solution of the $\mu$ problem is based on a mechanism originally
proposed \cite{dgp} in the context of gauge-mediated supersymmetry
breaking and can be viewed as a generalization of the pseudogoldstone
mechanism \cite{pseudogoldstone}. The crucial observation is that, 
whenever a large expectation
value of some higgs field is triggered through its couplings
with a gauge singlet field $S$, the supersymmetry breaking effects
shift the vev of $S$ from zero by an amount of the order of the low 
energy supersymmetry breaking scale
(which can be parameterized by the gravitino mass $m_{3/2} \sim 100\ 
{\rm GeV} - 1\ {\rm TeV}$ in theories with gravity-mediated 
supersymmetry breaking \cite{gravity-mediated}).
This allows one to induce the $\mu$ term 
via the coupling $Sh^{(1)}h^{(2)}$ in the superpotential, where 
$h^{(1)}, 
h^{(2)}$
denote the MSSM higgs doublets.
As we will see below, this shift of the vev of $S$ from zero is quite 
insensitive to the value of the large scale $M$. A priori, any 
symmetry group broken at $M$ is suitable for generating the
$\mu$ term. Our choice of $SU(2)_R\times U(1)_{B-L}$ is motivated
by the fact that it also can lead to 
`hot' dark matter in neutrinos with mass in the eV range.
 A global $U(1)$ R-symmetry
plays an essential role in our analysis. Its unbroken $Z_2$ subgroup 
acts
as `matter parity', which implies a stable LSP.  The R-symmetry
also gives rise to an accidental $U(1)_B$, including higher order terms 
in the superpotential, which leads to a stable proton. The model 
predicts
a spectral index of primordial density fluctuations that is extremely 
close to 
unity and which is consistent with a `cold' plus `hot' dark 
matter scenario of structure formation \cite{shst}.

Let us first describe the main features of the $G_{LR}$ symmetric model 
which 
solve the $\mu$ problem. In fact, the solution
can work in any scheme with gravity-mediated supersymmetry breaking 
provided there is an additional symmetry group factor broken
at some scale $>> m_{3/2}$ (or even simply a superheavy vev).
The $SU(2)_R\times U(1)_{B-L}$ group is broken by 
a pair of $SU(2)_R$ doublet chiral
superfields $l^c,\bar l^{c}$ which acquire a vev $M >> m_{3/2}$.
In order to achieve this breaking by a renormalizable superpotential, we 
will need a gauge singlet chiral superfield $S$. This singlet plays a 
crucial three-fold role: 1) it triggers $SU(2)_R$ breaking;
2) it generates  the $\mu$ and $B\mu$ terms of MSSM after supersymmetry 
breaking; and 3) it leads to hybrid inflation\cite{hybrid}. To see this, 
we first ignore the matter fields of the model and consider the 
superpotential
\begin{equation}
W = S( \kappa l^c\bar l^{c} + \lambda h^2 - \kappa M^2), \label{W1}
\end{equation}
where the chiral superfield $h = (h^{(1)}, h^{(2)})$ belongs to a 
bidoublet $(2,2)$ representation of
$SU(2)_L\times SU(2)_R\times U(1)_{B-L}$,
and $h^2$ denotes the unique bilinear invariant
$\epsilon^{ij}h_i^{(1)}h_j^{(2)}$. Note that, through a suitable 
redefinition
of the superfields, the parameters $\kappa, \lambda$ and $M$ can be made 
real and positive.
Also note that $W$ in Eq.(\ref{W1}) has the most general renormalizable 
form invariant under 
the
gauge group and
a continuous $U(1)$ R-symmetry  under which $S$ carries the same charge 
as $W$, while $h,l^c\bar l^{c}$ are neutral.
Clearly, a vev of $S$ will  generate a $\mu $ term with $\mu$ =$ \lambda 
<S>$.
Moreover, the vev of its F component, $F_S ={\partial W \over \partial 
S}$,
together with the soft trilinear gravity-mediated terms will generate a 
bilinear soft term,
$B\mu h^{(1)}\epsilon h^{(2)}$, in the scalar potential.

To understand how the $\mu$ problem is solved, let us analyze the 
minimum of the scalar potential.
 In the unbroken global supersymmetry limit, the vacuum is at
\begin{equation}
S = 0,~~~\kappa l^c\bar l^{c} + \lambda h^2 = \kappa M^2,~~~
l^c = e^{i\phi}\bar l^{c*}~~~h^{(1)}_i = 
e^{i\theta}\epsilon_{ij}h^{(2)j*},
\end{equation}
where the last two conditions arise from the requirement of D flatness.
We see that there is a flat direction (with two real dimensions)
at generic points of which
both $SU(2)_L$ and $SU(2)_R$ are spontaneously broken. The supersymmetry 
breaking will lift this degeneracy. Clearly, the desirable vacuum
is the one with $h^{(1),(2)} = 0$ and $l^c\bar l^{c} = M^2$
(up to higher order corrections).
Whether this indeed is the case depends on the parameters of the model
and the cosmological history (see below).

Let us investigate the
theory about this supersymmetric minimum. The $SU(2)_R$ is broken and 
all states
in $l^c, \bar l^{c}$ obtain masses $\sim M$ either through the 
superhiggs effect or through their mixing with the $S$ superfield. The 
masses of  $S$ and of the higgs components of $l^c,\bar l^{c}$ are 
$m_{infl}=m_S =m_{l^c}=\sqrt{2}\kappa M$. The only massless (up to 
supersymmetry breaking corrections) non-gauge degrees of freedom so far 
are
the two higgs doublets in $h$. Their `masslessness' can be simply 
understood
from their pseudogoldstone nature\cite{pseudogoldstone}: they are zero 
modes of the vacuum flat direction. Gravity-mediated soft terms lift the 
degeneracy along the
flat direction, the doublets acquire masses $\sim m_{3/2}$,
and the $\mu$ term is generated by a shift ($\sim m_{3/2}$) of the vev 
of
$S$ from zero.
Such an automatic generation of the $\mu^2 \sim B\mu (\sim m_{3/2}^2)$ 
terms after supersymmetry breaking is a generic feature of the models in 
which the higgs doublets are pseudogoldstone 
particles\cite{pseudogoldstone}.

To make the connection more transparent, let us take
$\lambda = \kappa$  for a moment. In this limit, the superpotential in 
Eq.(\ref{W1}) has an accidental $U(4)$ symmetry under which the $(l^c, 
h^{(1)})$
and $(\bar l^c,\epsilon h^{(2)})$ states transform as the fundamental 
and anti-fundamental representations respectively.
Of course, the $G_{LR}$
gauge interactions and the Yukawa couplings break the $U(4)$
symmetry explicitly, but this breaking cannot affect the vacuum 
degeneracy as long as
supersymmetry is unbroken. Thus, the degeneracy of the vacuum manifold
is as if we had an exact $U(4)$ symmetry broken to $U(3)$ by the vevs
of $l^c, \bar l^c$. This breaking produces seven would-be goldstone 
superfields, three of which are the true `goldstones' that are absorbed 
by 
the massive
$SU(2)_R\times U(1)_{B-L}/U(1)_Y$ gauge superfields. The remaining four 
are the physical states $h^{(1)},h^{(2)}$. They are `pseudogoldstones' 
and obtain masses only after supersymmetry breaking. At tree level, 
however, one combination, 
 $h^{(1)} +\epsilon h^{(2)*}$, must be exactly massless as a 
result of the $U(4)$ symmetry and only gets a mass by radiative 
corrections. This implies the generic 
relation $\mu^2 + m_{3/2}^2 =-B\mu$ in the $U(4)$ symmetric case.
{}For $\lambda \neq \kappa$, the $U(4)$ symmetry is explicitly broken 
in the superpotential and the above relation holds only approximately.
  
To see how all this works in detail, we write the full low-energy scalar 
potential
including the soft terms (for simplicity, we discuss the case with
a canonical K\"ahler metric, but the results 
stay essentially intact even in the general case). We have
\begin{eqnarray}
V= | \kappa l^c\bar l^{c} + \lambda h^2 - \kappa M^2|^2 +
(m_{3/2}^2 + \kappa^2 |\bar l^c|^2
+ \kappa^2 |l^c|^2 + \lambda^2 |h|^2)|S|^2 + m_{3/2}^2(|\bar l^c|^2 
\nonumber\\
+ |l^c|^2 + |h|^2) +  \left (Am_{3/2}S( \kappa l^c\bar l^{c} + \lambda 
h^2) 
- (A - 2)m_{3/2}S\kappa M^2 + {\rm h. c.} \right ) . \label{V}
\end{eqnarray}
Since $S$ and  the $l^c$,$\bar l^c$ fields are heavy in the
vanishing $m_{3/2}$ limit,
their vevs cannot be shifted significantly by the tiny soft 
supersymmetry breaking effects.
{}For a leading order estimate of the vev of $S$, we can substitute in $V$ 
the supersymmetric vevs of
the $SU(2)_R$ doublets. We see that $S$ gets a destabilizing tadpole 
term
$\simeq 2\kappa m_{3/2}M^2S + {\rm h.c.}$, and taking account of the 
term
$\simeq 2\kappa^2M^2|S|^2$, the resulting vev of $S$ 
is $\simeq - m_{3/2}/\kappa$. The vev of $S$ will  generate a $\mu$ term
with  $\mu =\lambda <S> = -m_{3/2} \lambda/\kappa$,
whereas the vev of its $F$-component
together with the soft trilinear gravity-mediated terms
will generate a $B\mu$ term in the scalar potential with
\begin{equation}
B\mu = \lambda \left ( F_S^* + m_{3/2}SA \right ). \label{BMU}
\end{equation}
The magnitude of $B\mu$ is readily
found using the equation of motion of $l^c$:
\begin{equation}
\kappa \bar l^c(\kappa l^c\bar l^{c} + \lambda h^2 - \kappa M^2)^* + 
(\kappa^2|S|^2 + m_{3/2}^2)l^{c*} +  Am_{3/2}S\kappa \bar l^c = 0.
\label{l^c}
\end{equation}
One obtains $B\mu \simeq -2\lambda m^2_{3/2}/\kappa$.
The above leading order estimates
can be confirmed by an explicit minimization of $V$ using the iterative 
series
$l^c = \bar l^{c} = M( 1 +  \sum_{n \geq 1} c_n (m_{3/2}/M)^n)$ and 
$S =-(m_{3/2}/\kappa)(1+ \sum_{n \geq 1} d_n(m_{3/2}/M)^n)$. 

 So far we were expanding the theory about the `good' minimum at 
$h=0,l^c=\bar l^c=M$, assuming that
this is the prefered ground state of the system. Now we must check the 
self-consistency of this assumption. One can show that, in the case of 
minimal K\"ahler potential, both the `good' ($h=0, l^c,\bar l^c \neq 0$) 
and the
`bad' ($h \neq 0, l^c,\bar l^c = 0$) points are local minima of the 
potential for all values of the parameters. In fact, they are the only 
minima for $ \kappa \neq \lambda$. For $\kappa = \lambda$, the 
degeneracy of the vacuum is not totally lifted and
these points lie on a flat direction, with one real dimension.
A simple way to see that the `good' stationary point is never unstable 
is to observe that the electroweak higgs mass squared matrix has no 
negative eigenvalues.  This is due to the fact that the diagonal
elements of this matrix are equal to $\lambda^2 |S|^2+m^2_{3/2}$, 
while its off diagonal elements are  $B\mu 
=(-\lambda/\kappa)(\kappa^2|S|^2+m^2_{3/2})$, as one can deduce from
Eqs.(\ref{BMU}) and (\ref{l^c}).  The determinant of this matrix is then 
equal to 
$(\lambda^2/\kappa^2-1)(\lambda^2 \kappa^2
|S|^4-m^4_{3/2}) $ which, replacing S by its leading order vev, becomes 
equal
to  $ (\lambda^2/\kappa^2-1)^2 m^4_{3/2} \geq 0 $.  This together with 
the obvious 
positivity of its trace imply that this matrix has no negative 
eigenvalues 
for all values of the parameters.  This statement remains true even if 
higher
order corrections to the vev of S are included.  
As it turns out, for any choice of the parameters, the `good' and `bad' 
minima of $V$ are degenerate to leading order (up to
corrections of order $m_{3/2}^4$ in the energy difference). If we allow
for a non-minimal K\"ahler potential, the degeneracy between these 
minima can be lifted.  In fact, we can find ranges of parameters where 
the `bad' point ceases even to be a local minimum.  For example, by 
replacing $(A-2)$ in Eq.(\ref{V}) by $(A-2\alpha)$, we can have the 
`good' point 
as the unique local minimum of the potential provided 
$|\alpha|^2 \geq \lambda/\kappa > 1$ or $|\alpha|^2 \leq \lambda/\kappa 
<1$. As we 
will see, the cosmological evolution prefers the first choice.

We are now ready to introduce the matter fields of the model as well.
The superpotential takes the following form:
\begin{equation}
W = S( \kappa l^c\bar l^{c} + \lambda h^2 - \kappa M^2)
+ g_{ab}^qhQ_aQ_b^c + g_{ab}^lhl_al^c_b + h_{ab}{\bar l^{c}\bar 
l^{c}l_a^cl_b^{c}\over M_P}, \label{W2}
\end{equation}
where $Q_a,Q_a^c, l_a,l^c_a$ are chiral quark, antiquark and lepton
superfields respectively and $a,b = 1,2,3$ are the family indices.
The last term gives superheavy masses to the right handed neutrinos 
which, through the seesaw mechanism \cite{sees}, can lead to a `tau' 
neutrino mass in the eV range.
The superpotential in Eq.(\ref{W2}) has the most general form (up to 
quartic terms) allowed by the gauge group and a 
$U(1)$ R-symmetry  under which the superfields have the following 
charges:
 $R_S = 1,~R_{l^c} = - R_{\bar{l^c}} = R_h = 0,
R_{Q_a} = R_{Q_a^c} = R_{l_a} = R_{l_a^c}= 1/2$. Note that the 
$Z_2$ subgroup of this R-symmetry remains unbroken and indeed is the 
conventional MSSM `matter parity'. 

The $U(1)$ R-symmetry, in contrast to `matter parity', also eliminates 
the dimension five\cite{proton}
operators responsible for proton decay.  If one assumes that the 
R-symmetry is an exact symmetry then, in fact, it eliminates
the baryon number violating operators from the superpotential to all 
orders.
Note that even if the R-symmetry is not explicitly broken by 
Planck scale
suppressed operators in the superpotential it must, in any case, be 
broken
by the hidden sector superfields which break supersymmetry spontaneously 
and also produce a non-vanishing vev of the superpotential (gravitino 
mass).
So, in general, we can expect some higher dimensional baryon number 
violating
terms coming from a nonminimal K\"ahler potential. These are, however, 
adequately suppressed.
The proton is
essentially stable in the present scheme.

Next let us  show that this model has a built-in inflationary trajectory
in the field space along which the $F_S$ term takes a constant value 
as in the supersymmetric hybrid inflationary scenario 
\cite{dss,lss}.
The relevant trajectory in the field space is parameterized by $S$.
The key point here is that $S$ has no self-interactions
and appears in the superpotential only linearly. At a generic point of 
this trajectory with $|S| > {\rm max \left ({M},
{M \sqrt{\kappa/\lambda}}\right )}$, all the gauge non-singlet higgs 
fields obtain masses of order 
$|S|$ and, therefore, they decouple. 
The massless degrees of freedom
along this trajectory are:
the singlet $S$, the massless $G_{LR}$ gauge supermultiplet,
and the massless matter superfields.
The effective low energy superpotential, which is obtained after 
integrating out the heavy superfields, can be readily constructed by 
simply 
using holomorphy and symmetry arguments.  This superpotential
is linear in $S$, namely
\begin{equation}
W_{infl} =-\kappa SM^2 .
\end{equation}
Were it not for the mass scale $M$ in the superpotential, the 
trajectory parameterized by $S$ would simply correspond to a 
supersymmetry preserving
vacuum direction remaining flat to all orders in perturbation theory.
The $F_S$ term, however, lifts this flat direction so that, at tree 
level,
 it takes an asymptotically constant value
for arbitrarily large $|S|$.
As a result, the trajectory of interest is represented by a massless
degree of freedom $S$, whose vev sets the mass scale for the heavy
particles and provides us with a constant tree level vacuum energy 
density
\begin{equation}
V_{{\rm tree}} = \kappa^2 M^4,
\label{dpot}
\end{equation}
which is responsible for inflation.

The above result can be easily
rederived by an explicit solution of the equations of motion along the 
inflationary trajectory \cite{dss,lss}. To this end, we can explicitly 
minimize all the D  and F terms for
large values of $|S|$. 
It is easy to check that, for
\begin{equation}
|S| > {\rm max \left ({M},
{M \sqrt{\kappa/\lambda}}\right)} , \label{cri}
\end{equation}
all the other vevs vanish and, therefore, a nonzero contribution to the
potential comes purely from the constant $F_S =- \kappa M^2$ term.
Whenever the condition in Eq.(\ref{cri}) is violated, the $l^c, \bar 
l^{c}$ components become tachyonic and
compensate the $F_S$ term. We see that, if $\kappa > \lambda$, $h$ will
become tachyonic earlier and the system will evolve towards the 
`wrong'
minimum. Thus, we prefer $\kappa < \lambda$. 
The  system rapidly approaches the supersymmetric
vacuum and oscillates about it. 
At tree level, the potential along the inflationary trajectory is
exactly flat \footnote{This statement is exact in the globally 
supersymmetric
limit as well as in supergravity provided the  K\"ahler metric for the 
field $S$ is canonical\cite{Hproblem}.
{}For generic K\"ahler potentials, however, this is no longer true.  For 
$|S|<<M_P$, the problematic term is $(SS^*)^2$,
whose coefficient must be somewhat smaller than unity.  Various 
inflationary regimes including this term are considered 
in refs.\cite{gr,linder,ddr}.}.
Radiative corrections, however, create a logarithmic slope \cite{dss} 
that drives the inflaton toward the minimum.
The origin of these corrections can be understood in the following way. 
As 
we have shown,
the value of $|S|$ sets the mass scale for the heavy particles along the
inflationary trajectory. Thus, we can think of the low energy theories 
at
different
points on this trajectory as being a single theory at different energy
scales. This gives rise to a wave function renormalization of the $S$ 
field
through loops involving the $\bar l^{c}, l^c, h$ particles.
Since their mass is set by the value of $|S|$, a nontrivial dependence 
on
this value arises providing an effective one-loop potential for the
inflaton field. For large field strengths or, in other words, for masses 
of the particles in the loop
suitably larger than  $M$,  this potential assumes the following form 
\cite{dss}
 \begin{equation}
\label{pot}
 V_{{\rm inf}} \simeq \kappa^2 M^4 \left [ 1 + {\kappa^2 +
\lambda^2 \over 8\pi^2}
{\rm ln}( {SS^* \over \Lambda^2} ) \right ].
\end{equation}
This simply is an asymptotic form of the one-loop
corrected effective potential\cite{colemanweinberg} with
\begin{equation}
\delta V_{one-loop} = {1 \over 64\pi^2} \sum_{i} (-1)^{F_i} 
M^4_i {\rm ln}( {M^2_i 
\over \Lambda^2} ). \label{ol}
\end{equation}
The contribution to Eq.(\ref{ol}) comes from the $\bar l^{c},l^c,~{\rm 
and}
~h$ supermultiplets, since 
they receive at tree level a non-supersymmetric contribution to the 
masses of
their scalar components from the $F_S$ term. All other states have 
either
no mass splitting due to a vanishing coupling with $S$ or have no 
inflaton dependent mass (these are the gauge and matter fields).
 
Inflation can end when the condition 
in Eq.(\ref{cri}) breaks down, thereby signaling that some of the fields 
become tachyonic and the system moves towards the
global minimum.  This is indeed the case provided the slow roll 
conditions
 \cite{dss} are not violated before this instability occurs.
As we have argued above,
for $\lambda > \kappa$, the system is destabilized towards the $SU(2)_R$
breaking vacuum and oscillates about it. These $S, l^c, \bar l^{c}$ 
oscillations
will create, among other particles, 
right handed neutrinos. The subsequent decay of these neutrinos gives 
rise to a primordial lepton number \cite{lepto}.  The observed baryon 
asymmetry of the universe can then be obtained by partial conversion of 
this lepton asymmetry through non-perturbative sphaleron effects 
\cite{krs}. This process, in
a certain range of the parameter space, can lead to a successful
baryogenesis via leptogenesis\cite{lepto}.  The gravitino constraint on 
the 
`reheat' temperature can also be simultaneously satisfied and the 
spectral 
index of primordial density fluctuations turns out to be extremely close
to unity.  The details are quite involved and will not be discussed 
here, but one fact is worth noting.  The $SU(2)_R \times U(1)_{B-L}$ 
breaking scale turns out to be about an order of magnitude lower than 
the MSSM unification scale and, thus, the right handed neutrino masses 
are restricted to be smaller than about $10^{13}$ GeV.  This implies 
that at least the `tau' neutrino mass can be in the eV range.

The model discussed above, although based on the $SU(2)_L\times SU(2)_R$
gauge group, is L-R asymmetric in the higgs sector. This is acceptable 
since, in reality, 
the above scheme must be a low energy remnant of a more fundamental 
theory in which the L-R
discrete symmetry is spontaneously broken at a scale 
$M_{LR} > M$.  We will now discuss some possibilities for such an 
embedding
of our model.
The superpotential can be generalized to be L-R symmetric 
 by adding a pair of chiral higgs $SU(2)_L \times U(1)_{B-L}$ 
doublet superfields $l,\bar{l}$:
\begin{equation}
W = S\left( \kappa (l^c\bar l^{c} + \bar ll) + \lambda h^2 - \kappa 
M^2\right ). \label{W}
\end{equation}
However, this structure is somewhat problematic.  The orientation of the 
superheavy vevs
in the L-R space is arbitrary due to the presence of an accidental 
global 
$U(4)$ symmetry in the
superpotential which leads to a larger continuous degeneracy of the 
vacuum
manifold. Therefore, at the point $l^c = \bar l^{c} = M$, the $l,\bar l$
doublet pair will appear exactly massless (in the supersymmetric limit) 
just as the $h$
superfield. Supersymmetry breaking effects will provide a small $(\sim 
M_W)$ mass to 
these doublets.  In order to maintain the gauge 
unification, we must supplement the low energy sector with some color 
triplets.
Note that this problem will not arise if the $l^c,\bar l^{c}$ 
and
$l, \bar l$ states are embedded in the $16, {\overline{16}}$ higgs 
representations
of the $SO(10)$ group so that the $l^c,\bar l^{c}$ vevs break $SO(10)$
down to $SU(5)$\cite{jeannerot}.
In this case, the $l, \bar{l}$ states will come in 
complete
multiplets under the $SU(5)$ subgroup and the unification of the gauge 
couplings will be unaffected. Moreover, all these additional states can 
get masses via the couplings of the $16$-plets
to the other higgs representations. We will not discuss the $SO(10)$
embedding in this paper.

In summary, we have presented a supersymmetric model based on the 
$SU(3)_c \times SU(2)_L \times SU(2)_R \times U(1)_{B-L}$ gauge group 
where the $\mu$ and  $B\mu$ terms are automatically generated after 
supersymmetry breaking. Moreover, the model gives 
rise to hybrid inflation.  The observed baryon 
asymmetry of the universe can be successfully generated through  
primordial leptogenesis, while the `hot' component of the dark matter in 
the universe consists of neutrinos with masses in the eV range. Finally, 
baryon 
number is essentially conserved.

\vspace{1.0cm}

We would like to thank J. Garcia-Bellido, S. King, R. Rattazzi and N. 
Tetradis 
for useful discussions.  One of us (Q.S.) would like to acknowledge the 
DOE support under grant number DE-FG02-91ER40626.
Q.S. and G.L. also acknowledge the NATO support, contract number
NATO CRG-970149. Finally, G.D. and G.L. acknowledge the TMR support 
under grant number ERBFMRXCT-960090.
 \begin{enumerate}

\end{enumerate}

\end{document}